\documentclass[fdp,11pt]{w-art}
\usepackage{times,cite,w-thm}   

\usepackage[latin1]{inputenc}
\usepackage{times}
\usepackage{amsfonts}
\usepackage{amssymb}
\usepackage{amsmath}

\def\xxFI{F^I}

\def\xxHi{H^i}

\def\xxGt{\mbox{$\tilde{G}$}}

\newcommand{\D}{\mathcal{D}}

\newcommand\cuadruplet{\mbox{$\mathbf{4}$} }
\newcommand\triplet{\mbox{$\mathbf{3}$} }
\newcommand\doublet{\mbox{$\mathbf{2}$} }
\newcommand\singlet{\mbox{$\mathbf{1}$} }

\usepackage{ifpdf}
\ifpdf
\usepackage[pdftex]{graphicx}
\usepackage[pdftex,unicode,implicit]{hyperref}
\hypersetup{%
  pdftitle    = {The General Gaugings of Maximal d9 Supergravity},
  pdfkeywords = {Supersymmetry, supergravity, symmetry, gauge },
  pdfauthor   = {J.J. Fernandez-Melgarejo, T. Ortin and E. Torrente-Lujan},
  pdfcreator  = {pdf\LaTeXe\ with package \flqq hyperref\frqq},
  pdfproducer = {pdf\LaTeXe\ with package \flqq hyperref\frqq},
  pdfpagemode = None,  
  pdffitwindow= true,  
  unicode     = true,
  plainpages  = true,
  colorlinks  = true,  
  citecolor   = blue,
  urlcolor    = red,
  linkcolor   = black
}
\newcommand{\hepth}[1]{arXiv:{\tt
\href{http://www.arXiv.org/abs/hep-th/#1}{hep-th/#1}}}

\newcommand{\arxiv}[1]{{\tt
\href{http://www.arXiv.org/abs/#1}{arXiv:#1}}}
\else
  \usepackage[dvips]{graphicx}
  \usepackage[unicode,implicit]{hyperref}
  \newcommand{\hepth}[1]{arXiv:{\tt hep-th/#1}}

  \newcommand{\arxiv}[1]{{\tt arXiv:#1}}
\fi

\begin{document}

\DOIsuffix{theDOIsuffix}
\Volume{65}
\Month{01}
\Year{2012}
\pagespan{1}{}
\Receiveddate{01/31/2012}
\Reviseddate{01/31/2012}
\Accepteddate{01/31/2012}
\Dateposted{01/31/2012}


\title[SUGRA 9D,Embedding Tensor]{Maximal Nine Dimensional Supergravity, General gaugings and the Embedding Tensor}


\author[ Fernandez-Melgarejo]{JJ. Fernandez-Melgarejo\inst{1}}
\author[ Ortin.]{T. Ortin\inst{2}} 
\author[ Torrente-Lujan]{E. Torrente-Lujan\inst{1,}
\footnote{Emails: jj.fernandezmelgarejo@um.es,Tomas.Ortin@csic.es,torrente@cern.ch.}}
\address[\inst{1}]{GFT, Dept. de Fisica, U. Murcia, Spain.}
\address[\inst{2}]{IFT UAM/CSIC
C/ Nicolas Cabrera, 13-15, C.U. Cantoblanco, E-28049-Madrid, Spain}

\begin{abstract}

We construct the most general maximal
    gauged/massive supergravity in $d=9$ dimensions and  determine its
    extended field content by using the embedding tensor method.


\end{abstract}
\maketitle

\section{Introduction}


The embedding-tensor formalism provides a systematic way of 
finding the extended field content of SUGRA Theories
(for recent reviews see
  Refs.~\cite{Trigiante:2007ki}
see also Refs.~\cite{Cordaro:1998tx,deWit:2002vt,deWit:2005hv,deWit:2005ub} and
\cite{deWit:2004nw,deWit:2007mt,Bergshoeff:2007vb,deWit:2008ta,Bergshoeff:2008bh,Hartong:2009az,Huebscher:2010ib}). 
One important feature
 of this formalism is that it requires the systematic
introduction of new higher-rank potentials which 
are related by St\"uckelberg
gauge transformations (that is,
 the \textit{tensor  hierarchy} of the theory
\cite{deWit:2005hv,deWit:2005ub,deWit:2008ta,Bergshoeff:2009ph,deWit:2009zv,Hartong:2009vc}). They 
can be taken as the (bosonic) extended field content of the theory. 
In Supergravity Theories one may need to take into account 
additional constraints on the possible gaugings, but, 
if the gauging is allowed  by supersymmetry,
then gauge invariance will require the introduction of 
all the fields in the associated tensor hierarchy and, 
since gauge invariance  is a pre- condition 
for supersymmetry, the tensor hierarchy will be automatically
compatible with supersymmetry. 
The  $d=9$ SUGRA   theory has (unless their 10D ascendants)
 three vector fields,  the embedding-tensor formalism 
is well suited to study all its possible gaugings and 
find its extended field content. 
Gaugings of the maximal $d=9$ supergravity have 
been obtained in the past
by generalized dimensional reduction 
\cite{Scherk:1979zr,Lavrinenko:1997qa}.
However, the possible combinations of
deformations were not studied and  some of the
higher-rank fields are associated to the constraints on the combinations of deformations. 
Furthermore, we do not know if other deformations, with no
higher-dimensional origin (such as Romans' massive deformation of the
$N=2A,d=10$ supergravity) are possible.
Our goal in this work will be to make a systematic study of all these
possibilities using the embedding-tensor formalism plus supersymmetry to
identify the extended-field content of the theory, 
finding the r\^ole played by the possible 7-, 8- and 9-form potentials, and compare the results with the
prediction of the $E_{11}$ approach. 

\section{Maximal $d=9$ supergravity: the undeformed theory}
\label{sec-undeformed}

There is only one undeformed (\textit{i.e.}~ungauged, massless) maximal
(\textit{i.e.}~$N=2$) 9-dimensional supergravity
\cite{Gates:1984kr}. 
The fundamental (\textit{electric}) fields of this theory are:  
($  e_{\mu}{}^{a},  \varphi, \tau\equiv  \chi+ie^{-\phi},A^{I}{}_{(1)}, B^{i}{}_{(2)}, C_{(3)}, \psi_{\mu}, \tilde{\lambda}, \lambda .$)
\footnote{
Where $I=0,\mathbf{i}$ ( indices of the 1- and 6-forms) with  $\mathbf{i,j,k}=1,2$ and
$i,j,k=1,2$ ( 2- and 5-forms). }
The complex scalar $\tau$ parametrizes an
$SL(2,\mathbb{R})/U(1)$ coset also described by a symmetric 
$SL(2,\mathbb{R})$ matrix $\mathcal{M}$  
(see \cite{article} for full definitions of these and other quantities).
Undeformed field strengths, $F^I,H^i,G$, of the electric $p$-forms 
are defined which are covariant
 under a set of abelian gauge transformations.
The field strengths follow a set of  simple Bianchi and 
(trivial)  Ricci identities ($dd F's=0$).
The bosonic action and the corresponding equations of motion can 
be  explicitly written. The latter  have as  global symmetry group
$G\equiv SL(2,\mathbb{R})\times (\mathbb{R}^{+})^{2}$. 
\footnote{
One of the  $(\mathbb{R}^{+})^{2}$ scaling transformations
 ($''\alpha''$)
acts on the metric and only leaves the
equations of motion invariant.  The second one, $''\beta''$, 
leaves invariant both the metric and the action: it is
 the \textit{trombone symmetry} which may
not survive to higher-derivative string corrections. }
The representation content of the theory is as follows.
Under  $SL(2,R)$ transformations: $( \phi,\tau,C_{(3)}):\singlet,B_{(2)}:
\doublet,A_{(1)}:\singlet+\doublet$.
The weights of the electric fields under all the scaling 
symmetries are given in   \cite{article}. 
The dimension of the global symmetry group,
which acts on the scalar manifold, 
is larger than that of the scalar manifold itself. 
There  is one Noether 1-form current $j_{A}$ associated to each 
of the generators of the global symmetries ($T_{A},A=1-5$) which 
are (under $SL(2,R)$) 
$j_A\sim (j_{\triplet},j_{\singlet},j_{\singlet^\prime})$. 
These currents are conserved on-shell.
 The SUSY rules are well known, to lowest order in fermions 
can be found in \cite{Bergshoeff:2002nv,Nishino:2002zi} 
(see \cite{article} for corrections).  
%

\vspace{0.2cm}
{\bf Equations of motion (EOMS)  and Magnetic fields}.
The EOMS for the electrical fields, 
can be written, after some algebra, in a particularly 
interesting form for our purposes of defining
 magnetic duals. They read:
\begin{align*}
d\left(e^{\frac{4}{\sqrt{7}}\varphi}\star F^{0} \right)& = 
-e^{-\frac{1}{\sqrt{7}}\varphi} \mathcal{M}^{-1}_{ij} 
F^{i} \wedge \star H^{j} +\tfrac{1}{2}G \wedge G\, ,\\  
d\left(e^{\frac{3}{\sqrt{7}}\varphi}\mathcal{M}^{-1}_{ij}
\star F^{j} \right)& = 
-e^{\frac{3}{\sqrt{7}}\varphi} \mathcal{M}^{-1}_{ij} 
F^{0} \wedge \star H^{j}
+ e^{\frac{2}{\sqrt{7}}\varphi}H_{i}\wedge \star G\, ,\\  
d\left(e^{-\frac{1}{\sqrt{7}}\varphi}\mathcal{M}^{-1}_{ij}\star H^{j} \right)
& =  e^{\frac{2}{\sqrt{7}}\varphi}F_{i} \wedge \star G
-H_{i}\wedge G\, ,\\
d\left(e^{\frac{2}{\sqrt{7}}\varphi}\star G \right)
& = F^{0} \wedge G+ \tfrac{1}{2}H^{i}\wedge H_{i}\, .
\end{align*}
 It is straightforward from these expressions to define the 
duality relations, directly for the magnetic  field strengths 
($ \tilde{G}_{(5)},\tilde{H}_{(6),i}, \tilde{F}_{(7),I}$) as:
\begin{align*}
\tilde{G}\, &\equiv e^{\frac{2}{\sqrt{7}}\varphi}  \star G, &
\tilde{F}_{0}& \equiv e^{\frac{4}{\sqrt{7}}\varphi}\star F^{0} ,
\\
 \tilde{H}_{i}& \equiv
e^{-\frac{1}{\sqrt{7}}\varphi}\mathcal{M}^{-1}_{ij}\star H^{j} ,&
\tilde{F}_{i}& \equiv e^{\frac{3}{\sqrt{7}}\varphi}\mathcal{M}^{-1}_{ij}\star F^{j} .
\end{align*}
The  relation of the magnetic dual  field strengths and dual 
potentials  ($  \tilde{C}_{(4)},\tilde{B}_{(5)}, \tilde{A}_{(6)}$) 
is not unique and will obtained afterwards.
Another, slightly different, way to define 
magnetic fields potentials and identify their field strengths,
 consists in writing the  equations of motion of the $p$-forms
as total derivatives. One can easily check that both methods 
lead to the same or compatible results.
We complete a full characterization of the dual
 fields  by recursion starting from the lowest levels of 
the hierarchy. We obtain in this way \cite{article}: 
their representation content under
 global transformations (weights, $SL(2,R)$ representations);
  explicit, hierarchy-compatible,  expressions for the 
gauge variations;
 explicit dual field strengths in terms of the potentials 
and their Susy transformations (possibly modulo dual relations). 
As usual, the Bianchi  identities  of the electric fields are the 
 EOMS for the Magnetic Duals while
the  EOMS of the Electric ones are the Bianchi identities 
 for the Magnetic fields.

\vspace{0.2cm}
{\bf Duals of scalars, and Noether currents.}
This dualization procedure is made possible by the gauge symmetries 
associated to all the $p$-form potentials for $p>0$ 
(massless $p$-forms with $p>0$).
For the scalars, 0-form fields, there is one Noether 1-form
current $j_{A}$ associated to each of the generators of 
the global symmetries of the theory $T_{A}$. 
These currents are conserved on-shell. 
We can define 
a $(d-2)$-form potential $\tilde{A}_{(d-2)}^{A}$ by
$d\tilde{A}_{(d-2)}^{A}= J^A\equiv  G^{AB}\star j_{B}.$
Thus, in summary,  the dualization procedure indicates that 
for each electric $p$-form with $p>0$ there is a 
dual magnetic $(7-p)$-form transforming in the
conjugate representation, there are as many magnetic 
$(d-2)$-form duals of the scalars as 
the dimension of the global group (and not of as the dimension 
of the scalar manifold) and that they transform in the
co-adjoint representation.

{\bf  \bf{N2}-SUGRA $D9$  boson tensor hierarchy. }
We write together  the   Bianchi identities for 
the electrical fields,  
the  dual magnetic fields (which are just the equations of 
motion of the electrical fields ) and 
the dual Noether currents (which are just the  
conservation equations for the   Noether forms) to get the 
( 1-8 rank ) hierachy of bosonic form equations which is 
shown in table\ref{tableforms} (left).
\begin{table}[h]
$$
\begin{array}{l|r}
\begin{array}{r}
d F_\phi     =  0 ,\quad d F_\tau    =  0 , \\  
d F^I     =  0 ,  \\  
d H^i   +  F^{0}  F^i  = 0 , \\
d G       -   F^i H_i = 0 , \\
d \tilde G   +  F^0 G + \frac{1}{2} \epsilon_{ij} H^i  H^j   = 0,   \\
d \tilde{H}_i  +  F_i  \tilde G -H_i  G= 0 , \\ 
d \tilde{F}_0   +     F^j \tilde{H}_j - \frac{1}{2}  G G = 0,  \\
d  \tilde{F}_i  +  F^0  \tilde{H}_i - H_i  \tilde G = 0,\\
d J_A  =0.
\end{array}
&
\begin{array}{r}
\D \vartheta =0,\quad  \D Z's  = 0,\\
 \D F_\phi     = Z_{\phi,I} F^I ,\quad \D F_\tau   = Z_{\tau,I} F^I, \\  
\D F^I     =  Z^I_i H^i, \\  
\D H^i    +  F^{0}  F^i   = Z^i G, \\
\D G      -  F^i H_i = Z \tilde{G}, \\
\D \xxGt   + 
 F^0 G +\frac{1}{2} H^i  H_i   = Z^i H_i,   \\
\D \tilde{H}_i  +  F_i  \xxGt  - H_i  G  = Z_i^I \tilde{F}_I^0,\\
\D\tilde{F}_0 +F^i\tilde{H}_i-\frac{1}{2} G G=\vartheta_0^A J_A,\\
\D \tilde{F}_i   +   F^0  \tilde{H}_i - H_i \xxGt = \vartheta_i^A J_A.\\
\end{array}
\end{array}
$$
\caption{(Left) Undeformed Bianchi hierarchy.
  We have also added the 
trivial Bianchi equations corresponding to the 
1-form ``field strengths'' (not related by duality):
$ F_\phi\equiv d\phi, F_\tau\equiv d\tau$. 
(Right) Deformed Bianchi Hierarchy.
We also add  the conditions which lead to
 the gauge invariance of the (constant) deformation parameters. 
Indices are lowered and raised using the 
antisymmetric tensor $\epsilon_{ij}$ where appropiate. }
\label{tableforms}
\end{table}

\section{Deforming the maximal $D9$ supergravity}
The  maximal 9D SUGRA  has  3$\times$  1-forms $A^I$ potentials 
which can be  used  to gauge  some  unspecified ``part'' of  the 
global    $G=SL(2,R)\times R^2$ symmetry acting on 
the scalar sector.
We promote the constant symmetry parameters to local ones
$\alpha^{A} \to \Lambda^{I}(x)\vartheta_{I}{}^{A}$,
where: $\Lambda^{I}(x)$ are  gauge parameters and  
 $\vartheta\equiv\vartheta_{I}{}^{A}$ is the  
{\em embedding tensor}
   \cite{Cordaro:1998tx,deWit:2002vt,Bergshoeff:2009ph}.
We require the theory to be invariant under the local transformations
$\delta_{\Lambda}\varphi =\Lambda^{I}\vartheta_{I}{}^{A}k_{A}{}^{\varphi},$
$\delta_{\Lambda}\tau =\Lambda^{I}\vartheta_{I}{}^{A}k_{A}{}^{\tau}$.
It is useful to  define a covariant derivative using the 
embedding tensor.
For scalars and, in general any $(r)_G$-form, we define
$\displaystyle \D \eta^{(r)}= d\eta^{(r)}+ \delta_\Lambda( \eta^{(r)})=d\eta^{(r)}+ A^J X_J (\eta^{(r)})$.
 Where   $ X_{J}^{(r)} =\vartheta_J^A T^{(r)}_{A}$, a linear combination
of  the 
generators $T_A$ in the representation $(r)$ of $G$
We request that the covariant 
derivative follows a  Leibnitz rule 
$\D(XY)\sim \D X Y+\epsilon_p X \D Y$.
This a non-trivial condition because in principle
the  $X_J$'s do not  follow any Jacobi identity.

Thus, we are going to require invariance under the new 
gauge transformations for the scalar fields, we will 
 find that we need new couplings to the gauge 1-form fields 
(as usual). Repeating this procedure on the p-forms 
we can see we need the  coupling to the $(p+1)$-forms, etc.
One explicitly sees that the derivatives  of the scalars
and field strengths of any rank are covariant if we consistently
assume modified gauge variations for $p$-form potentials
where $(p+1)$-potentials are included as Stueckelberg terms.
Most of these fields are already present in the 
supergravity theory or can be identified with their magnetic duals 
but this procedure allows us to introduce
consistently the highest-rank fields 
(the $d$-, $(d-1)$- and $(d-2)$-form potentials), which are not 
dual to any of the original electric fields. 
Actually,  the highest-rank potentials are related to the 
symmetries (Noether currents), 
the independent deformation parameters 
and the constraints that they satisfy.

A general result of the embedding tensor formalism tell us 
that, similarly, we need  to introduce p-form potentials in the 
expressions of the field strengths.
The field strengths will follow modified Bianchi and Ricci identities.
The procedure of requesting gauge invariance becomes equivalent 
to the study of a tensor hierarchy of form potentials and strengths 
and their associated identities (the Bianchi and Ricci ones) and
 their possible deformations.
Requesting gauge invariance amounts to the overall compatibility
of the deformed Bianchi and Ricci hierarchies with the result of 
the appearence of strong conditions on any deformation parameters.
We write in table\ref{tableforms}(right)  the summary of the 
deformed Bianchi  hierarchy.
This deformation basically consist of: 
the substitution of the standard derivatives by newly 
defined ones and by adding Stuelckelberg terms (``Z'' terms). 
The Bianchi hierarchy is supplemented  by a hierarchy of
Ricci-like identities, which are of the form:
$$\D\D\xxFI   = X_{(JK)}{}^I F^J\wedge F^K , \D\D\xxHi   = X_{Jk}{}^i F^J\wedge H^k ,...\D\D F_{(n)} = X_{(n)} F_{(2)}\wedge F_{(n-2)}  .$$

The initial bosonic  deformations parameters  
of the theory are (24 in total):  
$\vartheta_I^A,Z^i,Z_i^I,Z$.
We note that if the embedding tensor
 $\vartheta\to 0$ then, both, the covariant 
derivative and Stueckelberg couplings  $\D\to d$, $Z\to 0$, 
we recover the undeformed theory. 
But, on the opposite,
If we  take $Z's\to 0$ then 
(as can be checked by direct computation) the embedding tensor
  $\vartheta\to 0$ (and then $\D\to d$, so we 
are dealing  only with gauging deformations).
No new degrees of freedom are introduced.
It is implicitly assumed that the global symmetry group $G$ is 
unbroken. 
No other massive deformation parameters, beyond $\vartheta$ and the
$Z's$ are permitted,
this is what apparently SUSY prefers 
\footnote{This is requested if we want to keep the modifications of the
fermion SUSY rules to a minimum.
Other deformation tensors will request, at least, drastic SUSY deformations.}.
Finally we note the role played by the singlet $Z$ parameter.
This parameter acts as a switch of the coupling of 
the electrical ($p=1,2,3$) and magnetic ($p>3$) 
field strengths.
We see that, if $Z\not =0$, we are in presence of 
a twisted/derivative self-duality type condition 
($\star G\sim \frac{1}{Z} \D G+..$) and we are lead to a 
non-action theory.   
We will require that any deformation parameter is 
gauge invariant, this is trivially equivalent to 
impose the conditions $\D\theta=\D Z's=0$
\footnote{These conditions can be seen as the zero-degree
identities in the Bianchi hierarchy}.

The gauge invariance of the theory is equivalent to the overall 
consistency of the Bianchi and Ricci hierarchies and
 imposes strong conditions on the deformations 
parameters $\vartheta,Z's$.
The general structure  of the compatibility equations between 
the Bianchi and Ricci hierarchies can be outlined.
For any n-field strength $F_{(n)}$, we have:
\begin{eqnarray*}
 \D F_{(n)}+R_{(n)}&=& Z_{(n)} F_{(n+1)},
\quad\quad \D\D F_{(n)}=X_{(n)} F_{(2)} F_{(n)},
\end{eqnarray*}
applying a covariant derivative and using $\D Z_{(n)}=0$ we get
$\D\D F_{(n)}+D R_{(n)}= Z_{(n)} \D F_{(n+1)}.$
Using the assumed Ricci and Bianchi indentities 
and imposing the Leibnitz Rule for a derivation,  we get 
a set of algebraic identities  of the form
\footnote{where
$ R_{(n)} = c_n F_{(2)} F_{(n-1)}+...,$,
$D R_{(n)} = c_n F_{(2)} DF_{(n-1)}+...= c_n Z_{(n-1)}  F_{(2)} F_{(n)}-c_n F_{(2)} R_{(n-1)}+...$.}
\begin{eqnarray*}
X_{(n)} F_{(2)} F_{(n)}+
 c_n Z_{(n-1)} F_{(2)} 
F_{(n)} + c_{n+1} Z_{(n)} F_{(2)} F_{(n)}+...
&=& Z_{(n)} Z_{(n+1)}  F_{(n+2)}.
\end{eqnarray*}
As a final result, we get  a set of linear 
(of the type $X_{(n)} + c_n  Z_{(n-1)}  + c_{n+1} Z_{(n)} =0$)
and cuadratic constraints ($Z_{(n)} Z_{(n+1)}  =0$)
involving the deformation parameters. 
The latter are, in detail,
$ Z_i{}^I \vartheta_I{} ^A =  Z^i Z_i{} ^I  =  Z Z^i   = 0 . $

Furtherly, the  gauge invariance of the, constant, 
deformation parameters ( summarized by  $\D\vartheta=0,\D Z's=0$) 
imply
 a  new set of conditions.  The 
first one, the {\em   cuadratic constraint}
\begin{eqnarray}
\vartheta X_J^{(ad)} -  X_J^{(3)} \vartheta &=&0, 
\label{eq:cuadratic}
\end{eqnarray}
and similar conditions of type
$Z_{(n)} X_J^{(n+1)} - X_J^{(n)}  Z_{(n+1)} =0$.

We get explicit  expressions for the deformed 
field strengths in terms of 
the potentials and the gauge variations of the potentials 
by recursion from the lowest degree, assuming the compatibility 
of the hierarchy.

We proceed  to the deformation of the SUSY rules.
At any rank, we  study the modifications of the supersymmetry 
transformation rules of the scalars and fermion fields which
are needed to ensure the closure of the local supersymmetry 
algebra. 
We replace the derivative and field strengths appearing 
in the fermion SUSY rules by the new covariant ones and 
add some new ``deformation'' parameters, the 
 \textit{fermion  shifts}  ($f,k,g,h,\tilde{g},\tilde{h}$).
No derivatives or field strengths appear at the bosonic 
 SUSY rules.
Moreover we do not introduce any other massive deformation parameter 
at these bosonic form SUSY rules. This is consistent with keeping
to a minimum the deformation of the bosonic tensor hierarchy 
(where we introduced only the $\vartheta,Z's$ parameters).
The    commutators of any two  SUSY 
transformation are expressions involving all the 
deformation parameters ($\vartheta,Z's$, fermion shifts)
\footnote{ i.e. for one of the scalars 
$\left[\delta_{\epsilon_{1}},\delta_{\epsilon_{2}}\right]\varphi= 
\xi^{\mu}\partial_{\mu}\varphi +(..
\Re{\rm e}(\tilde{h}),\Im{\rm m}(\tilde{g}),\Re{\rm e}(\tilde{g})..).$}.
We require that these deformed commutators 
 can consistently be written in the generic form
$ [\delta,\delta]=\delta_{gct}+\delta_{\Lambda}+(duality)$
to lowest order in fermions.
We have explicitly checked that, at any rank,
 we can consistenly find values for the
general coordinate transformation 
and gauge transformations ($\Lambda$) so that the 
above general form of the commutators is obtained.
The procedure allow us to find
 expressions for the 
fermion shifts in terms of the other deformation parameters 
(ultimately in terms of only the embedding tensor) and the 
scalars of the theory
 using the  bosonic gauge constraints and a new linear constraint
  $X_1^{(\tau)}\tau+X_2^{(\tau)}=0$
\footnote{A similar relation in appear in the application of 
the embedding tensor formalism in 4D SUGRA  $\vartheta_i^AP_A=0$}. 
Moreover no duality relations are needed (closure off-shell)
\footnote{At least up degree 4}.

\section{Reduction of parameters.}  
The effect of the linear constraints is twofold:
in first place allow us to express all the deformation parameters
in terms of only  the embedding tensor, 
in second place  reduces the number of independent components of this 
embedding tensor. These components, moreover, can be assigned to
definite $G=SL(2,R)\times R^2$ representations.
From a total of 24 initial parameters 
we are left in a first stage with   8 Parameters
( $\vartheta_{\singlet},\vartheta_{\doublet},\vartheta_{\doublet},\vartheta_{\triplet}$).
\footnote{
The identification is as follows:
$\vartheta_{\singlet} \equiv \vartheta_0^5 \sim Z  (\sim m_{IIB})$,
$\vartheta_{\doublet} \equiv \vartheta_i^4 \sim (Z^i\pm  Z^0_i) (\sim (m_4,\tilde{m}_4))$,
$\vartheta_{\doublet} \equiv \vartheta_i^5 \sim (Z^i\pm Z^0_i) (\sim (m_{11},m_{IIA}))$,
$\vartheta_{\triplet} \equiv \vartheta_0^m \sim Z^i_j (\sim (m_2\pm m_3,m_1)).$}
The representation content (weights) with respect 
scaling transformations is given  in \cite{article}.
Let us study the effect of the cuadratic constraints.
The  ``ZZ''-type   contraints
( $Z_J^A\vartheta_I^A=Z_j^IZ^j=Z Z^j=0$) are automatically 
satisfied after imposing the previous linear 
constraints and the  cuadratic constraint given by 
Eq.(\ref{eq:cuadratic}).
This cuadratic constraint decomposes in $G$-irreducible 
representations as follows 
($Q_{\cuadruplet}, 4\times Q_{\doublet},Q_{\singlet}$) (see  \cite{article}.
The compatibility of these set of non-linear constraints additionally 
imposes other secondary constraints.

The identification of the minimal set of ``irreducible''
cuadratic constraints is a subtle issue.
Let us study in some detail an interesting subcase. 
The 
 $\vartheta_{\singlet}(\sim Z\sim m_{IIB})$ 
component of the embedding tensor 
controls the coupling of the $G-\tilde{G}$  Bianchi 
identities (see the bosonic tensor hierarchy).
%
If we set  $\vartheta_{\singlet}=0$ (or $Z=0$) and we
fix  generic, non trivial, values for 
the other components ($\vartheta_{\triplet},\vartheta_{\doublet}\not=0$) 
the set of independent cuadratic constraints reduces to:
{
\begin{equation*}
Q_{\cuadruplet} : \left(\vartheta_{0}{}^{m}
\left(12\vartheta_{i}{}^{4}+5\vartheta_{i}{}^{5} \right)\right)_{\cuadruplet}=0,\
Q_{\doublet^{\prime\prime}}: 
\vartheta_{j}{}^{4} \left(\vartheta_{0}^{m} T_{m}
\right)_{i}{}^{j}  = 0\, ,\
Q_{\doublet^{\prime\prime\prime}}: \vartheta_{j}{}^{5} \left(\vartheta_{0}^{m} T_{m}\right)_{i}{}^{j} = 0.
\end{equation*}
}
The compatibility of these constraints  implies the following
secondary constraints:
$\det (T_m \vartheta_0^m)=\vartheta_0^{1,2}+ \vartheta_0^{2,2}- \vartheta_0^{3,2} =0,\quad  \vartheta^4_i = \lambda \vartheta^5_i$,
which can be seen as reducing the number of effective parameters.
We see that: 
a) the list of constraints  reduces to  
$ Q_{\cuadruplet},Q_{\doublet^{\prime\prime}},Q_{\doublet^{\prime\prime\prime}}$, in accordance 
with $E_{11}$ predicitions,
b) the naive counting of effective parameters is 
(number parameters-number constraints)
$\sharp=(3+2+2)-(1+1)=3+2$.
We are left with a (non-linear) triplet and a doublet of parameters,
they are not the originals but non linear functions of them.
The conclusion is that 
if $\vartheta_{\singlet}=0$ the
embedding tensor  simply agree with 
$\sim$ $E_{11}$ predictions~ \cite{Bergshoeff:2010xc,Bergshoeff:2011zk}
\footnote{The dimensional reduction of
the adjoint representation of the $E_{11}$ algebra  predicts the
$SL(2,R)\times R^+$ $D9$ field content,
      (p=7): $\sim(\triplet,\doublet)$, 
      (p=8): ($\triplet,\doublet$),
      (p=9): $(\cuadruplet,\doublet,\doublet)$.
}.

\section{Conclusions.}
In this work  we have applied the embedding-tensor formalism to 
the study of
the most general deformations (\textit{i.e.}~gaugings and massive
deformations) of maximal 9-dimensional supergravity. 
We have used the complete global $SL(2,\mathbb{R})\times \mathbb{R}^{2}$ symmetry of its equations of motion. 
We have found the constraints on the deformation parameters imposed by 
gauge and SUSY invariance
 (the latter imposed through
the closure of the local supersymmetry algebra to lowest order in
fermions) . 
The minimal set of deformation parameters 
(8=$\triplet+\doublet+\doublet+\singlet$) 
appears in agreement  with 
Ref.~ \cite{Bergshoeff:2002nv}.
 We have  found explicit expressions for the 
   field strengths, gauge  and
SUSY sules  of the deformed theory.
According to the general embedding tensor framework the 
possible extra 7-, 8- and 9-forms,  are
respectively dual to the Noether currents, 
independent deformation tensors and irreducible quadratic constraints. 
Thus all the higher-rank fields have an interpretation in terms
of symmetries and gaugings. 
We can conclude that we have satisfactorily identified 
the extended field
content (the tensor hierarchy) of maximal 9-dimensional 
supergravity and,
furthermore, that all the higher-rank fields have an interpretation in terms
of symmetries and gaugings. 
 In  comparison with the $E_{11}$ level decomposition  
 \cite{Bergshoeff:2011zk}, when comparable, the ETF gives similar 
results. 

\vspace{0.3cm}
{\small
{\bf Acknowledgments. }
TO would like to thank E.~Bergshoeff for several useful conversations.  
This
work has been supported in part by the Spanish grants FPA2009-07692,
FIS2007-1234, FPA2008-453, HEPHACOS S2009ESP-1473, CPAN-CSD2007-00042,
MU-Seneca-2007/3761.  
The work of JJFM has been supported by the Spanish 
MEC FPU-AP2008-00919 grant.  
JJFM and ET would like to thank CERN and IFT-UAM/CSIC for their
hospitality.
TO wishes to thank
M.M.~Fern\'andez for her permanent support.
}

\end{document}